\documentclass[prx,aps,superscriptaddress,footinbib,twocolumn]{revtex4-1}
\usepackage{mathrsfs}
\usepackage{amsfonts}
\usepackage{amssymb}
\usepackage{graphicx}
\usepackage[usenames,dvipsnames]{color}
\usepackage[colorlinks=true,citecolor=blue,linkcolor=magenta]{hyperref}
\usepackage{lmodern}
\usepackage{dsfont}
\usepackage{natbib}
\usepackage{amsmath}
\usepackage{bm}
\usepackage{pgf}
\usepackage{calligra}
\usepackage{algorithmicx}
\usepackage{amsmath} 
\usepackage{amsmath, braket}
\usepackage{booktabs}
\usepackage{multirow}
\usepackage{microtype}
\usepackage[caption=false]{subfig}
\usepackage{graphicx}
\usepackage{subfig}
\usepackage{overpic}
\usepackage{rotating}

\begin{document}

\title{Quantum computing for effective nuclear lattice model}

\author{Zhushuo Liu}
\affiliation{Graduate School of China Academy of Engineering Physics, Beijing 100193, China}

\author{Jia-ai Shi}
\affiliation{Graduate School of China Academy of Engineering Physics, Beijing 100193, China}

\author{Bing-Nan Lu}
\email{bnlv@gscaep.ac.cn}
\affiliation{Graduate School of China Academy of Engineering Physics, Beijing 100193, China}

\author{Xiaosi Xu}
\email{xsxu@gscaep.ac.cn}
\affiliation{Graduate School of China Academy of Engineering Physics, Beijing 100193, China}

\begin{abstract}
Nuclear lattice effective field theory has become an important framework for quantum many-body calculations in nuclear physics, yet its classical implementation remains increasingly challenging for more general interactions and larger systems. In this work, we develop a quantum-computing framework for a three-dimensional nuclear lattice model. We construct a variational quantum eigensolver framework and systematically compare the Jordan--Wigner and Gray code encodings. Our analysis shows that for the few-body systems considered here, Gray code combined with symmetry reduction yields a substantially more compact qubit representation. Based on this framework, we perform numerical studies for $^{2}\mathrm{H}$, $^{3}\mathrm{H}$, and $^{4}\mathrm{He}$ on finite lattices. The calculated ground-state energies exhibit a clear approach toward the corresponding experimental binding energies as the lattice size increases. These results provide a proof-of-principle foundation for future quantum simulations of nuclear many-body problems.
\end{abstract}

\maketitle

\section{Introduction}

Quantum many-body problems are commonly addressed by discretizing spacetime on a lattice, most often a cubic one.
This approach underlies several important frameworks, including lattice quantum chromodynamics in high-energy physics~\cite{PhysRevD.10.2445,RevModPhys.55.775},
nuclear lattice effective field theory (NLEFT) in nuclear physics~\cite{PPNP63-117}, and lattice simulations for cold-atom systems~\cite{RevModPhys.80.885}. A lattice provides a natural discrete representation of many-body quantum states, while the continuum limit can in principle be approached by extrapolating to zero lattice spacing and infinite volume~\cite{RevModPhys.51.659,Lahde:2019npb}. 
On classical computers, such lattice models are usually studied with many-body numerical methods such as exact diagonalization and quantum Monte Carlo (QMC)~\cite{Fehske_2008}. 
In favorable cases where QMC is free of the sign problem, the computational cost can scale polynomially~\cite{PRL135-222504}. 
However, such sign-problem-free situations are limited, and for generic interactions the sign problem leads to an exponential growth of computational cost with system size, which severely restricts the accessible simulation range~\cite{PhysRevLett.94.170201,PhysRevB.41.9301}.

This difficulty is particularly relevant for nuclear lattice calculations. NLEFT has become an important \textit{ab initio} framework for nuclear structure calculations~\cite{PPNP63-117,Springer2019}, and has achieved many notable results for light and medium-mass nuclei~\cite{PhysRevLett.132.062501,Elhatisari_2015,Elhatisari2024}. At the same time, extending lattice nuclear calculations to more general interactions, higher precision, and more challenging nuclear systems remains computationally demanding, as the sign problem can lead to a rapid growth of statistical noise and computational cost. This motivates exploring alternative approaches, especially in regimes where classical methods become expensive.

Quantum computing offers a potentially different way to represent and manipulate many-body wave functions~\cite{PRXQuantum.4.027001}, and may help overcome some of the limitations of classical approaches for strongly correlated quantum systems. In nuclear physics, since the pioneering cloud quantum computation of the deuteron~\cite{PhysRevLett.120.210501}, existing quantum-algorithm studies have focused mainly on the shell model~\cite{perezobiol2023a,PhysRevC.105.064317,PhysRevC.106.034325,PhysRevC.105.064308,Bhoy2024}, while comparatively little attention has been paid to lattice formulations and nuclear effective field theories~\cite{gqdydvps,watson2025}. Yet nuclear lattice Hamiltonians are especially suitable for quantum simulation, since the lattice formulation already provides a finite-dimensional many-body Hilbert space that can be mapped directly to qubits. The main obstacle is the choice of representation. In a three-dimensional lattice model with four spin–isospin flavors per site, a direct Jordan–Wigner (JW) mapping~\cite{1928ZPhy47631J} requires $4L^3$ qubits. This scaling is compatible with fault-tolerant devices but becomes demanding on near-term hardware at moderate $L$. Since larger lattices are needed to reduce finite-volume effects, this qubit overhead quickly becomes prohibitive even for few-body systems on near-term devices.

For the few-body regime considered here, the key issue is therefore how to represent the symmetry-constrained low-energy Hilbert space more efficiently. A possible route is to first reduce the effective Hilbert space using symmetries and then encode only the reduced basis~\cite{Sandvik_2010}. Gray code encoding~\cite{PhysRevA.103.042405,PhysRevC.104.034301,bbkffjxj} follows this approach and maps the symmetry-adapted basis states of the reduced Hilbert space directly to a compact qubit register.

From the algorithmic perspective, quantum phase estimation~\cite{kitaev1995,Cleve1998} and quantum signal processing~\cite{PhysRevLett.118.010501,PRXQuantum.2.040203} provide rigorous frameworks for eigenvalue problems in principle, but they typically require deep circuits and substantial quantum resources. 
In the noisy intermediate-scale quantum (NISQ) regime~\cite{NISQ,RevModPhys.94.015004}, variational methods are more practical. 
Among them, the variational quantum eigensolver (VQE)~\cite{Peruzzo_2014,tilly2022VQE,cerezo2021VQE} uses parameterized quantum circuits of relatively low depth and has been widely applied to many-body problems in 
quantum chemistry~\cite{Grimsley2019,Kandala2017,Cao2019} and 
condensed-matter physics~\cite{PhysRevA.92.042303,PhysRevB.102.235122,Alvertis2025}. 
This makes VQE a natural starting point for exploring lattice nuclear Hamiltonians on quantum devices.

In this work, we study a quantum-computing approach to nuclear lattice problems from this perspective. Rather than attempting a full NLEFT calculation with realistic chiral interactions, we consider a simplified three-dimensional lattice Hamiltonian with an exact kinetic term and contact two-body and three-body interactions. This model retains the essential structure of a nontrivial nuclear lattice problem while allowing us to investigate the quantum representation of the Hamiltonian, the symmetry reduction of the many-body Hilbert space, and the performance of variational calculations for few-body systems. We construct the Hamiltonian in second quantization and compare the direct JW encoding with a Gray code encoding built on a symmetry-reduced basis. For the systems studied here, with particle numbers $n=2,3,4$ and lattice sizes $L=2$--$6$, symmetry reduction combined with Gray code encoding yields a much more compact qubit representation than the direct JW mapping. Based on this framework, we perform VQE calculations for $^{2}\mathrm{H}$, $^{3}\mathrm{H}$, and $^{4}\mathrm{He}$ on finite lattices.

The remainder of this paper is organized as follows. 
Section~II introduces the lattice Hamiltonian studied in this work. 
Section~III presents the VQE framework and the two qubit encodings. 
Section~IV describes the symmetry reductions used to reduce the effective Hilbert space and the corresponding qubit requirements. 
Section~V presents numerical results, and Section~VI concludes the paper.

\section{The nuclear lattice model}

In NLEFT, nuclear interactions, often derived from chiral effective field theory~\cite{RMP81-1773,PhysRept503-1}, are regularized on a cubic lattice, and many-body observables are computed in the resulting discretized Hilbert space. 
For very small systems, the problem can be solved with exact diagonalization methods such as Lanczos iteration, while for larger systems auxiliary-field quantum Monte Carlo (AFQMC) is typically employed. 
Over the past decade, NLEFT has been successfully applied to a broad range of problems, including 
ground states~\cite{PhysRevLett.104.142501,PLB732-110,PhysRevLett.128.242501,Elhatisari2024,PRL135-222504},
excited states~\cite{PhysRevLett.112.102501,Nat.Comm.14-2777},
nuclear densities~\cite{PhysRevLett.109.252501,PhysRevLett.119.222505}, 
scattering observables~\cite{PhysRevC.86.034003,Nature528-111}, and 
finite-density nuclear matter~\cite{PLB850-138463,PhysRevLett.132.232502}. 
Despite these advances, extending lattice nuclear calculations to more general and more realistic interactions remains highly challenging, largely because of the sign problem. 
This motivates the search for alternative computational approaches, including quantum algorithms.

Our goal is not to reproduce the full NLEFT framework in the present work. 
Instead, we consider a simplified nuclear lattice Hamiltonian that captures the basic structure of few-body lattice calculations while remaining suitable for a systematic study of quantum encodings, symmetry reductions, and variational algorithms. 
Specifically, we study a model with an exact kinetic operator and onsite contact two-body and three-body interactions. 
This provides a controlled starting point for investigating how lattice nuclear Hamiltonians can be represented and solved on a quantum computer.

\subsection{Lattice Hamiltonian}

Throughout this work, we consider a nuclear Hamiltonian for fermions with four spin-isospin degrees of freedom on a three-dimensional cubic lattice with periodic boundary conditions. 
In first quantization, the Hamiltonian is written as
\begin{equation}\label{H_1}
\begin{split}
    H &= \sum_{\alpha = 1}^{n} \frac{\mathbf{p}_{\alpha}^{2}}{2m} 
    + \sum_{\alpha<\beta}^{n} V_{2}(\mathbf{r}_{\alpha}, \mathbf{r}_{\beta}) \\
      &\quad + \sum_{\alpha<\beta<\gamma}^{n} V_{3}(\mathbf{r}_{\alpha}, \mathbf{r}_{\beta}, \mathbf{r}_{\gamma}),
\end{split}
\end{equation}
where $n$ is the particle number and $m$ is the nucleon mass. 
The first term is the kinetic energy, while the second and third terms represent two-body and three-body interactions, respectively. 
Higher-body interactions are neglected in the present model. 
Here $\mathbf{p}_\alpha=-i\nabla_\alpha$ is the momentum operator of particle $\alpha$, and $\mathbf{r}_\alpha$ is its position. 
The interactions are taken to be contact terms,
$$
V_2 = c_2\,\delta(\mathbf{r}_\alpha-\mathbf{r}_\beta), 
\qquad
V_3 = c_3\,\delta(\mathbf{r}_\alpha-\mathbf{r}_\beta)\delta(\mathbf{r}_\beta-\mathbf{r}_\gamma),
$$
with coupling constants $c_2$ and $c_3$.

After discretizing space and applying second quantization, Eq.~\eqref{H_1} becomes
\begin{equation}\label{H_2}
\begin{split}
    H &= \sum_{\sigma=0}^3 \sum_{l,l'=0}^{L^3-1} T_{ll'} \hat{c}_{l,\sigma}^{\dagger}\hat{c}_{l',\sigma}
    + c_2 \sum_{\sigma<\sigma'}^3 \sum_{l=0}^{L^3-1} \hat{n}_{l,\sigma}\hat{n}_{l,\sigma'} \\
    &\quad + c_3 \sum_{\sigma<\sigma'<\sigma''}^3 \sum_{l=0}^{L^3-1} \hat{n}_{l,\sigma}\hat{n}_{l,\sigma'}\hat{n}_{l,\sigma''},
\end{split}
\end{equation}
where $L^3$ is the total number of lattice sites and $T_{ll'}$ is the hopping matrix element between sites $l$ and $l'$. 
The kinetic term is obtained by Fourier transforming the continuum kinetic operator, which leads to an exact lattice representation with all-to-all hopping rather than a nearest-neighbor truncation. 
The index $\sigma\in\{0,1,2,3\}$ labels the four spin-isospin states
\[
\{p\uparrow,\; p\downarrow,\; n\uparrow,\; n\downarrow\},
\]
where $p$ ($n$) denotes proton (neutron) and $\uparrow$ ($\downarrow$) denotes spin up (spin down). 
The operators $\hat{c}_{l,\sigma}^{\dagger}$ and $\hat{c}_{l,\sigma}$ create and annihilate a fermion of flavor $\sigma$ at site $l$, and
\[
\hat{n}_{l,\sigma}=\hat{c}_{l,\sigma}^{\dagger}\hat{c}_{l,\sigma}
\]
is the corresponding number operator.

In this work, we use the parameter values
\[
c_2 = -143.202887, \qquad c_3 = 107.232753,
\]
together with lattice spacing $a=1.160747~\mathrm{fm}$. 
The couplings are fixed at $L=6$ by fitting $c_2$ to the deuteron binding energy and $c_3$ to the triton binding energy. 
The resulting parameters are then used to compute $^{4}\mathrm{He}$ as a consistency check, and are kept fixed when the lattice size $L$ is varied.

\section{VQE Framework for Nuclear Hamiltonians}

In this section, we provide a brief introduction to the variational quantum eigensolver (VQE), which is employed to compute the ground-state energy of the nuclear Hamiltonian. The VQE algorithm relies on the Rayleigh-Ritz variational principle~\cite{tilly2022VQE,cerezo2021VQE}. According to this principle, the expectation value of the Hamiltonian $H$ with respect to a parameterized trial state $|\psi(\boldsymbol{\theta})\rangle$ provides an upper bound on the true ground-state energy $E_g$:
\begin{equation}
    E_g \le \frac{\langle \psi(\boldsymbol{\theta})| H | \psi(\boldsymbol{\theta}) \rangle}{\langle \psi(\boldsymbol{\theta}) | \psi(\boldsymbol{\theta}) \rangle}.
\end{equation}

A typical VQE workflow consists of the following steps.

\noindent \textbf{Hamiltonian representation}: The first step in VQE is to construct the target Hamiltonian in a form suitable for quantum computation.

\noindent \textbf{Encoding}: By employing a fermion-to-qubit encoding scheme, such as the Jordan--Wigner transformation, the fermionic Hamiltonian is mapped to a qubit Hamiltonian $H=\sum_i c_i P_i$. Here, $P_i \in \{I, X, Y, Z\}^{\otimes N}$ represents an $N$-qubit Pauli string that can be directly measured on a quantum computer.

\noindent \textbf{Ansatz construction}: An expressive yet depth-efficient parameterized quantum circuit $U(\boldsymbol{\theta})$ is constructed to prepare the trial state $|\psi(\boldsymbol{\theta})\rangle=U(\boldsymbol{\theta})|\phi_0\rangle$, where $|\phi_0\rangle$ is the initial state.

\noindent \textbf{Measurement}:~The cost function $E(\boldsymbol{\theta})=\langle\psi(\boldsymbol{\theta})|H|\psi(\boldsymbol{\theta})\rangle$ is estimated by Hamiltonian averaging using finite-shot measurements. To reduce measurement overhead, commuting Pauli terms can be grouped and measured simultaneously.

\noindent \textbf{Optimization}: The parameters $\boldsymbol{\theta}$ of the ansatz are updated iteratively using a classical optimizer (e.g., COBYLA, Adam, SPSA or SLSQP) until the energy converges.

In the following, we provide a more detailed explanation of our choice of encoding and ansatz tailored to the nuclear lattice model considered in this work.

\subsection{Encoding}

Implementing quantum algorithms for nuclear Hamiltonian eigenproblems necessitates a well-defined encoding to map fermionic basis states and operators onto their qubit counterparts~\cite{PhysRevC.104.034301,PhysRevA.103.042405}. The choice of encoding strategy critically governs the computational resource overhead, specifically the qubit count, circuit depth, and gate complexity.
In this study, we evaluate two distinct schemes: the canonical Jordan--Wigner transformation and the Gray code encoding. The two approaches differ fundamentally in their utilization of physical symmetries and the efficiency with which they represent the relevant Hilbert space.

\subsubsection{Jordan--Wigner Transformation}

Among the available fermion-to-qubit encoding schemes, the Jordan--Wigner (JW) transformation is the most straightforward and widely adopted. To account for the four spin-isospin degrees of freedom inherent to nuclear systems, the qubit register is partitioned into four sectors, each corresponding to a distinct spin-isospin state.
Each sector comprises $L^3$ qubits, matching the number of lattice sites.
Within each block, the qubit state $|1\rangle_j$ ($|0\rangle_j$) signifies that the $j$-th lattice site is occupied (unoccupied) by a fermion. The fermionic creation and annihilation operators are mapped to qubit operators as follows:
\begin{align}
  \hat{c}_j^\dagger &= \frac{1}{2}(X_j - i Y_j) \otimes \bigotimes_{k<j} Z_k, \\
  \hat{c}_j &= \frac{1}{2}(X_j + i Y_j) \otimes \bigotimes_{k<j} Z_k,
\end{align}
where the nonlocal string of $Z$ operators preserves the fermionic anti-commutation relations and ensures the correct anti-symmetry of the wave function.

For the Hamiltonian in Eq.~\eqref{H_2}, the number of hopping terms scales as $\mathcal{O}(L^6)$, while the on-site multi-body terms scale as $\mathcal{O}(L^3)$. After the JW mapping, the number of Pauli terms is still dominated by the hopping part and scales as $\mathcal{O}(L^6)$ up to a constant factor.

While the JW encoding is conceptually straightforward, it poses significant challenges for noisy intermediate-scale quantum (NISQ) devices. Achieving high-precision results typically requires a large lattice size $L$, resulting in qubit overheads that exceed the capacity of current hardware. Moreover, the physically relevant Hilbert subspace---constrained by particle number conservation and other symmetries---is exponentially smaller than the full qubit space spanned by the JW encoding. This discrepancy motivates the adoption of compact encoding schemes  (for example, the Gray code encoding introduced in the next section) that can exploit symmetries to truncate the Hilbert space. In the era of fault-tolerant quantum computing, the JW encoding may regain prominence; its resource requirements scale linearly with the lattice volume rather than the particle number, offering a scalable pathway once qubit limitations are alleviated.

\subsubsection{Gray Code Encoding}

To minimize qubit overhead, we first construct a symmetry-adapted basis that respects all relevant symmetries, such as particle-number conservation and momentum conservation (as detailed in the subsequent section). This procedure constrains the problem to a reduced Hilbert subspace of dimension $N$.

Subsequently, we employ the Gray code encoding to map each physical basis state $\ket{k}$ in this subspace to a multi-qubit state $\ket{g_k}$, where $g_k \in G_\eta$.
Here, $\eta = \lceil \log_2 N \rceil$ represents the minimum number of qubits required, and $G_\eta$ denotes the set of $\eta$-bit binary strings ordered such that adjacent elements have a Hamming distance of exactly one. This dense encoding scheme effectively utilizes the qubit Hilbert space, resulting in an exponential reduction in qubit requirements compared to the Jordan--Wigner encoding.

For instance, consider a reduced subspace of dimension $N=8$; this requires only $\eta = \log_2(8) = 3$ qubits. The corresponding Gray code mapping is given by:
\begin{align*}
    \ket{0} &\mapsto \ket{000}, & \ket{1} &\mapsto \ket{001}, & \ket{2} &\mapsto \ket{011}, & \ket{3} &\mapsto \ket{010}, \\
    \ket{4} &\mapsto \ket{110}, & \ket{5} &\mapsto \ket{111}, & \ket{6} &\mapsto \ket{101}, & \ket{7} &\mapsto \ket{100}.
\end{align*}

Due to the lack of a direct physical interpretation for the Gray code basis states, the fermionic creation and annihilation operators cannot be straightforwardly mapped to compact local qubit operations. Consequently, the qubit Hamiltonian $H_q$ is constructed by explicitly projecting the Hamiltonian matrix elements onto the encoded basis states:
\begin{equation}
    H_q = \sum_{i, j=0}^{N-1} \langle i | H | j \rangle \ket{g_i}\bra{g_j},
\end{equation}
where $N$ denotes the dimension of the reduced Hilbert space. The outer products $\ket{g_i}\bra{g_j}$ are then decomposed into tensor products of single-qubit operators using the following relations:
\begin{align*}
    \ket{0}\bra{1} &= \sigma^{+} = \frac{1}{2}(X + iY), &
    \ket{1}\bra{0} &= \sigma^{-} = \frac{1}{2}(X - iY), \\
    \ket{0}\bra{0} & = \frac{1}{2}(I + Z), &
    \ket{1}\bra{1} & = \frac{1}{2}(I - Z).
\end{align*}

As the qubit Hamiltonian is built from matrix elements in the reduced basis, after expanding the operators into Pauli strings, the number of terms can grow quickly with the reduced dimension $N$; in the worst case it scales as $\mathcal{O}(N^2)$ (up to cancellations). For this reason, Gray code encoding is mainly useful when the reduced Hilbert space remains small, which in our setting corresponds to the low-particle-number regime.

\subsection{Hardware-Efficient Ansatz}

Since the Gray code basis lacks a direct physical interpretation, it prevents the straightforward application of physically motivated ansatzes such as unitary coupled cluster (UCC). However, because the encoding itself enforces all relevant symmetries, there is no need to impose additional symmetry constraints at the circuit level. This significantly reduces the required gate overhead. For these reasons, we employ the hardware-efficient ansatz (HEA) in this work. The HEA consists of repeated layers of parameterized single-qubit rotations and two-qubit entangling gates. The structure makes it well suited for near-term quantum devices.

\begin{figure}[h]
    \centering
    \includegraphics[width=1\linewidth]{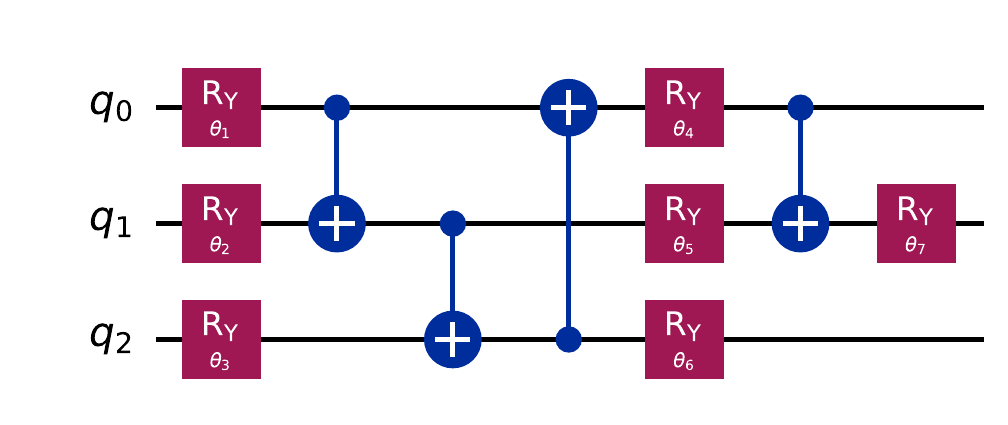}
    \caption{Schematic diagram of a 3-qubit streamlined hardware-efficient ansatz.}
    \label{HEA}
\end{figure}

Since the ground-state wave function of the nuclear Hamiltonian is real-valued, we adopt a simplified ansatz in which the single-qubit gates are restricted to $R_y$ gates, which is sufficient to represent real amplitudes while keeping the circuit depth minimal. A schematic diagram of a 3-qubit ansatz is shown in Fig.~\ref{HEA}.

\section{Quantum resources reduction using symmetries}

\subsection{Dimension of the Symmetrized Hilbert Space}

We start from the nuclear Hamiltonian defined in Eq.~\eqref{H_2}. The model is formulated on a three-dimensional cubic lattice with periodic boundary conditions. At each lattice site there is a single spatial orbital and four internal degrees of freedom corresponding to spin and isospin. Therefore, each site can host at most four fermions due to the Pauli principle.

Throughout this work, we consider a fixed total particle number $n\ll4L^3$. The full Fock-space dimension at fixed particle number is given by the number of ways to distribute $n$ fermions among the available single-particle states. Since the total number of single-particle states is $4L^3$, the dimension of the Hilbert space is

\begin{equation}
    \dim(\mathcal{H})  = \binom{4L^3}{n}.
\end{equation}

If proton and neutron numbers are fixed separately, $N_p$ and $N_n$, then the dimension becomes
\begin{equation}
    \dim(\mathcal{H})  = \binom{2L^3}{N_p}\binom{2L^3}{N_n}.
\end{equation}

We now use the spatial symmetry group $G$ to construct projection operators and restrict the Hilbert space to the subspace consistent with the symmetry of the ground state. For a finite cubic lattice with periodic boundary conditions, this symmetry group $G$ is given by the semidirect product of the discrete translation group $T$ and the cubic point group $O_h$:

\begin{equation}
    G = T \rtimes O_h,
\end{equation}
where $T$ is the finite Abelian group formed by $L^3$ discrete translations, and $O_h$ denotes the full cubic point group comprising 48 elements (including rotations, reflections, and spatial inversion). In practice, we treat the translation and point group symmetries separately.

First, for a translationally invariant Hamiltonian, the ground state lies in the zero-momentum sector ($\mathbf{k}=\mathbf{0}$). we therefore use the translation group $T$ to construct the projection operator onto this subspace:
\begin{equation}
    \hat P_{k=0} = \frac{1}{|T|}\sum_{\mathbf{t} \in T} \hat{T}_\mathbf{t},
\end{equation}
where $|T|=L^3$ is the number of lattice translations. Given an arbitrary Fock state $\ket{\psi}$ with fixed particle number, a corresponding zero-momentum state can be constructed as
\begin{equation}
    \ket{\Psi_{k=0}} = \frac{1}{\sqrt{|T|}}\sum_{\mathbf{t} \in T} \hat{T}_\mathbf{t} \ket{\psi},
\end{equation}
where the overall normalization depends on the orbit size of
$|\psi\rangle$ under the translation group.
For generic configurations, the orbit has size $|T|$, so the
normalization reduces to $1/\sqrt{|T|}$. In general, by exploiting translational invariance, we reduces the Hilbert space dimension approximately by a factor of $L^3$, up to corrections from configurations that are invariant under certain translations.

Next, we reduce the Hilbert space dimension by using point group symmetries. In the zero-momentum sector ($\mathbf{k}=\mathbf{0}$), the wave vector is invariant under any point group operation $R$ (i.e., $R \,\mathbf{0} = \mathbf{0}$). Consequently, the associated little group $G_{\mathbf{k}}$ coincides with the full cubic point group $O_h$.  
For a translationally invariant system without spontaneous symmetry breaking, a non-degenerate ground state transforms according to the trivial irreducible representation of the lattice symmetry group. In the cubic group $O_h$, this representation is $A_{1g}$, which corresponds to a lattice scalar. The corresponding projection operator is thus defined as:

\begin{equation}
    \hat{P}_{A_{1g}} = \frac{1}{|O_h|} \sum_{R \in O_h} \chi^{(A_{1g})}(R)^* \hat{R} = \frac{1}{48} \sum_{R \in O_h} \hat{R},
\end{equation}
where the character $\chi^{(A_{1g})}(R) = 1$ for all $R \in O_h$. The symmetry-adapted basis state $\ket{\Phi_{\text{sy}}}$ is explicitly derived as:
\begin{equation}
    \ket{\Phi_{\text{sy}}} = \hat{P}_{A_{1g}} \ket{\Psi_{k=0}} = \frac{1}{\sqrt{48 |T|}} \sum_{R \in O_h} \sum_{\mathbf{t} \in T} \hat{R} \hat{T}_{\mathbf{t}} \ket{\psi}.
\end{equation}

Ultimately, the cumulative application of these spatial symmetries yields a substantial reduction of the size of the Hilbert space relevant for the ground state. Starting from the Fock space with fixed particle number $n$, whose dimension is $\binom{4L^3}{n}$, projection onto the zero-momentum and $A_{1g}$ sectors reduces the dimension by approximately a factor of $48L^3$. Consequently, the dimension of the resulting symmetry-reduced Hilbert space can be estimated by:

\begin{equation}
    \dim(\mathcal{H}) \approx \frac{1}{48 L^3}\binom{4L^3}{n}.
\end{equation}
In this work, we consider $n \le 4$ and $L \in \{2, 3, 4, 5, 6\}$. In the regime where $L^3 \gg n$, the binomial coefficient can be approximated as:
\begin{equation}
     \binom{4L^3}{n}\approx \frac{(4L^3)^n}{n!}.
\end{equation}
which leads to:
\begin{equation}
    \dim(\mathcal{H}) \approx \frac{4^n(L^3)^{n-1}}{48 n!}.
\end{equation}

If proton and neutron numbers are fixed separately, the Fock-space dimension becomes $\binom{2L^3}{N_p}\binom{2L^3}{N_n}$. After imposing translation and point-group symmetries, the reduced dimension is approximately:
\begin{equation}
    \dim(\mathcal{H}) \approx \frac{2^n(L^3)^{n-1}}{48 N_p!N_n!}.
\end{equation}

\subsection{Symmetry-Based Qubit Reduction}

In this section, we use Gray code encoding to map the symmetry-adapted basis states. As discussed above, encoding an $N$-dimensional Hilbert space requires a qubit register size of $\lceil \log_2 N \rceil$. Consequently, with the spatial symmetry reduction, the required number of qubits $N_q$ is approximated by:

\begin{equation}
    N_q \approx 2n+ (n-1) \log _2(L^3) - \log_2 n! - \log_2 48.
\end{equation}

For comparison, the JW encoding maps each lattice site and internal degree of freedom directly to qubits. In that case, the required number of qubits scales as $N_q^{\text{JW}} = 4L^3$.  Note that one may taper a few qubits using $\mathbb{Z}_2$ parity symmetries associated with conserved particle numbers (and, when applicable, spin-resolved particle numbers). This reduces the register by a constant number of qubits but does not change the $\mathcal{O}(L^3)$ scaling.

Table~\ref{qubit-comparison} compares the Gray code encoding with the JW encoding for different lattice sizes $L$. In this calculation, we fix $N_p=1$, $N_n=2$ and total spin $S=1/2$, in addition to the spatial symmetry constraints discussed above. The qubit numbers reported in the table are obtained under these symmetry restrictions.

The JW encoding scales with the lattice volume $L^3$, while the Gray code encoding scales approximately with $\rm{log}_2(L^3)$ after symmetry reduction. As $L$ increases from 2 to 6, the qubit number required by JW grows from 24 to 648. In contrast, the Gray code encoding requires 9 qubits at $L=6$. 

This comparison shows that symmetry reduction combined with Gray code encoding significantly lowers the qubit requirement in the low-particle-number regime considered here. However, since the encoded basis no longer has a direct physical interpretation, physics-inspired ansatz constructions are not readily available, and the resulting variational circuits typically require greater depth for a fixed qubit number.

\begin{table}[h]
\centering
\caption{Comparison of the qubit count for Gray Code (GC) and Jordan--Wigner (JW) encodings as a function of the linear dimension $L$ of the three-dimensional lattice, at a particle number $n=3$.}
\label{qubit-comparison}
\setlength{\tabcolsep}{9pt} 
\renewcommand{\arraystretch}{1.2} 
\begin{tabular}{c c c}
\toprule[1.2pt] 
\multirow{2}{*}{Linear Dimension($L$)} & \multicolumn{2}{c}{Qubit Count ($N_q$)} \\ 
\cmidrule(lr){2-3} 
 & Gray Code & Jordan--Wigner  \\
\hline 
2 & 3 & 24 \\
3 & 4 & 81 \\
4 & 6 & 192 \\
5 & 7 & 375 \\
6 & 9 & 648 \\
\bottomrule[1.2pt] 
\end{tabular}
\end{table}

\section{Numerical simulations}

\begin{figure*}[htp]
    \centering
    \subfloat[Deuteron ($^{2}\mathrm{H}$)]{
        \includegraphics[width=0.32\linewidth]{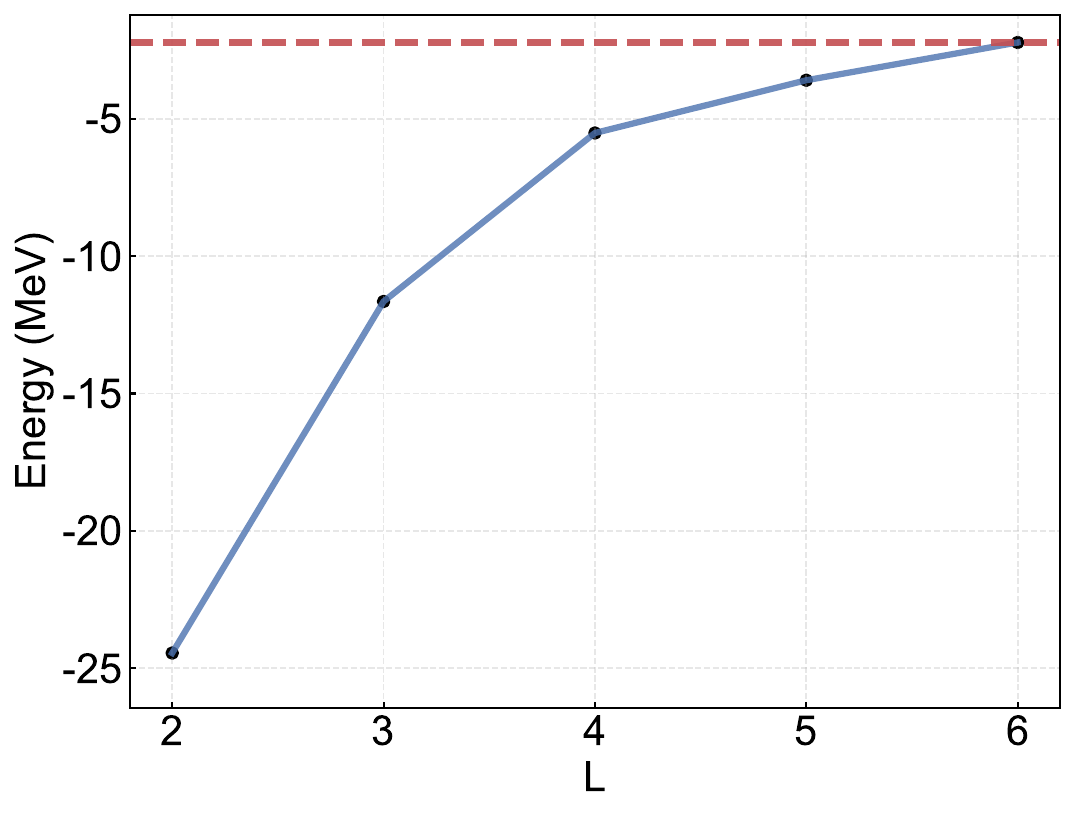} \label{N2E}
    }
    \hfill
    \subfloat[Triton ($^{3}\mathrm{H}$)]{
        \includegraphics[width=0.32\linewidth]{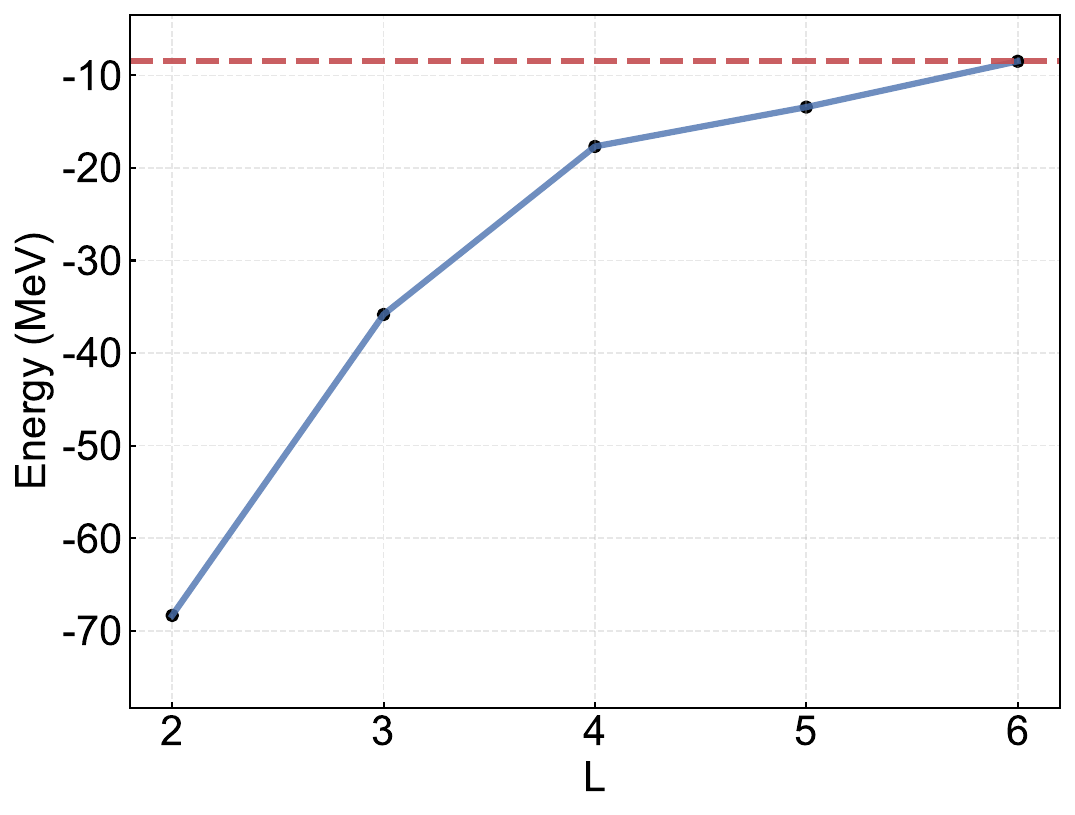} \label{N3E}
    }
    \hfill
    \subfloat[Helium-4 ($^{4}\mathrm{He}$)]{
        \includegraphics[width=0.32\linewidth]{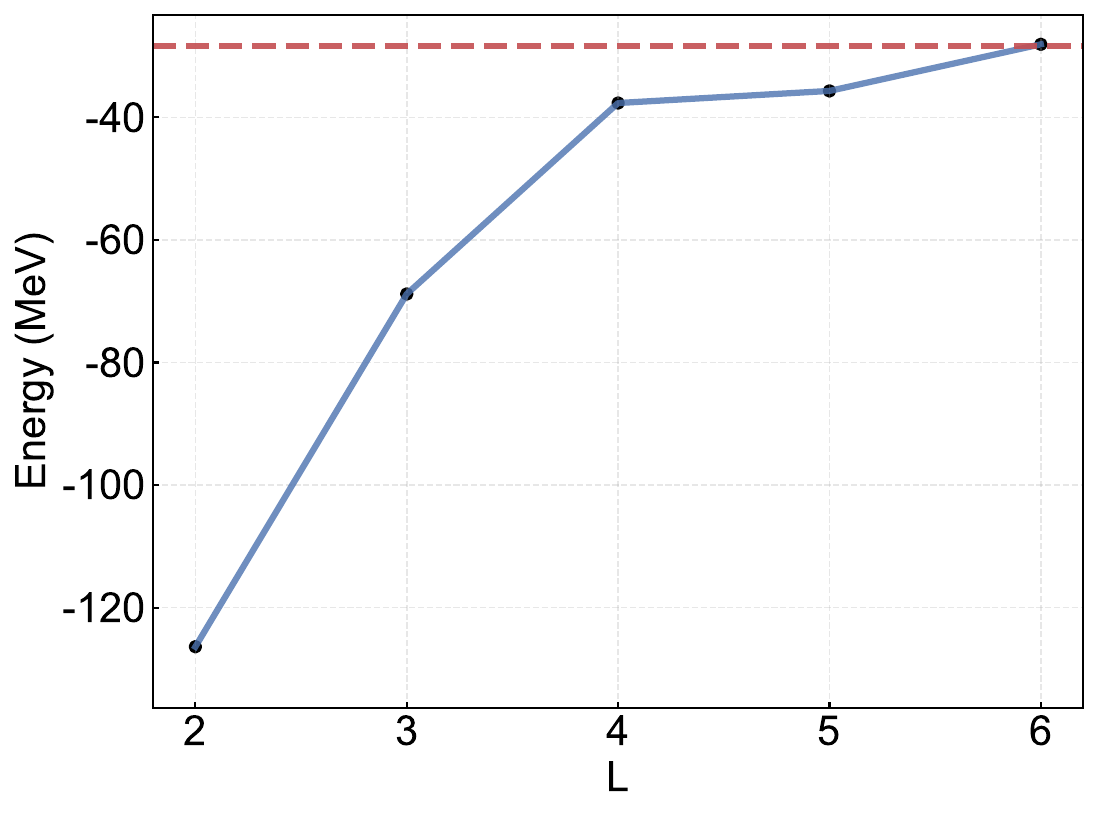} \label{N4E}
    }
    \caption{Ground-state binding energies versus lattice size $L$. Each data point (black circle) represents the variational energy minimum achieved via VQE optimization for the corresponding lattice dimension $L$ and nucleon number $n$. Panels display results for (a) the deuteron ($^{2}\mathrm{H}$, $n=2$), (b) the triton ($^{3}\mathrm{H}$, $n=3$), and (c) helium-4 ($^{4}\mathrm{He}$, $n=4$). Red dashed lines indicate the experimental binding energy benchmarks ($E_{\text{exp}}^{^{2}\mathrm{H}} \approx -2.22~\mathrm{MeV}$, $E_{\text{exp}}^{^{3}\mathrm{H}} \approx -8.48~\mathrm{MeV}$, and $E_{\text{exp}}^{^{4}\mathrm{He}} \approx -28.30~\mathrm{MeV}$). 
    }
    \label{NE}
\end{figure*}

In this section, we present a numerical investigation of the ground-state properties of light nuclei, including deuteron($^2$H), triton($^3$H), and helium-4($^4$He) using VQE. For each nucleus, the calculation is performed in the
corresponding fixed particle-number sector, with proton and
neutron numbers chosen according to the physical system. Additional symmetry constraints, including spatial
symmetries and conserved spin quantum numbers when
applicable, are imposed in the construction of the reduced
Hilbert space. The nuclear systems are modeled on finite lattices with linear dimension ranging from $L=2$ to $L=6$. Due to the prohibitive qubit overhead associated with the standard Jordan-Wigner transformation at these lattice sizes, we exclusively employ the resource-efficient Gray code encoding to map the nuclear states onto the qubit computational basis. All variational protocols are emulated classically using the PennyLane framework~\cite{bergholm2022pennylane}.
Gradients are computed efficiently via hardware-accelerated automatic differentiation using JAX~\cite{jax2018github}, and the variational parameters are optimized employing the Adam and L-BFGS-B algorithms implemented in the Optax library~\cite{deepmind2020jax}.

Figure~\ref{NE} shows how the ground-state binding energies depend on $L$ for $^{2}\mathrm{H}$, $^{3}\mathrm{H}$, and $^{4}\mathrm{He}$. Our main purpose here is to examine the finite-volume trend
of the calculated energies as the lattice size varied.

We first fix the interaction parameters at $L=6$. At this lattice size, we fit the two-body coupling strength $c_2$ in Eq.~\eqref{H_2} so that the calculated ground-state energy of $^{2}\mathrm{H}$ matches the experimental binding energy. We then fit the three-body coupling $c_3$ using the experimental ground-state energy of $^{3}\mathrm{H}$. After this fit, we keep the same values of $(c_2,c_3)$ for all other lattice sizes. With these fitted parameters, we compute the ground-state energy of $^{4}\mathrm{He}$ as a check. As shown in Fig.~\ref{N4E}, the $^{4}\mathrm{He}$ result at $L=6$ is also very close to the experimental value. In practice, if heavier systems are considered, the same fitted $(c_2,c_3)$ will be used.

When a smaller $L$ is adopted, the results deviate away from the experimental binding energies for all three nuclei. The deviation is largest at small $L$. This behavior is mainly a finite-volume effect under periodic boundary conditions: when the lattice is small, the wave function overlaps more strongly with its periodic images, and the long-range part of the bound-state wave function is distorted by the limited lattice size. As $L$ increases, these effects become smaller, and the energies approach the experimental benchmarks.

\begin{figure*}[t!] 
    \centering

    \subfloat[Deuteron ($^2$H)]{
        \includegraphics[width=0.32\linewidth]{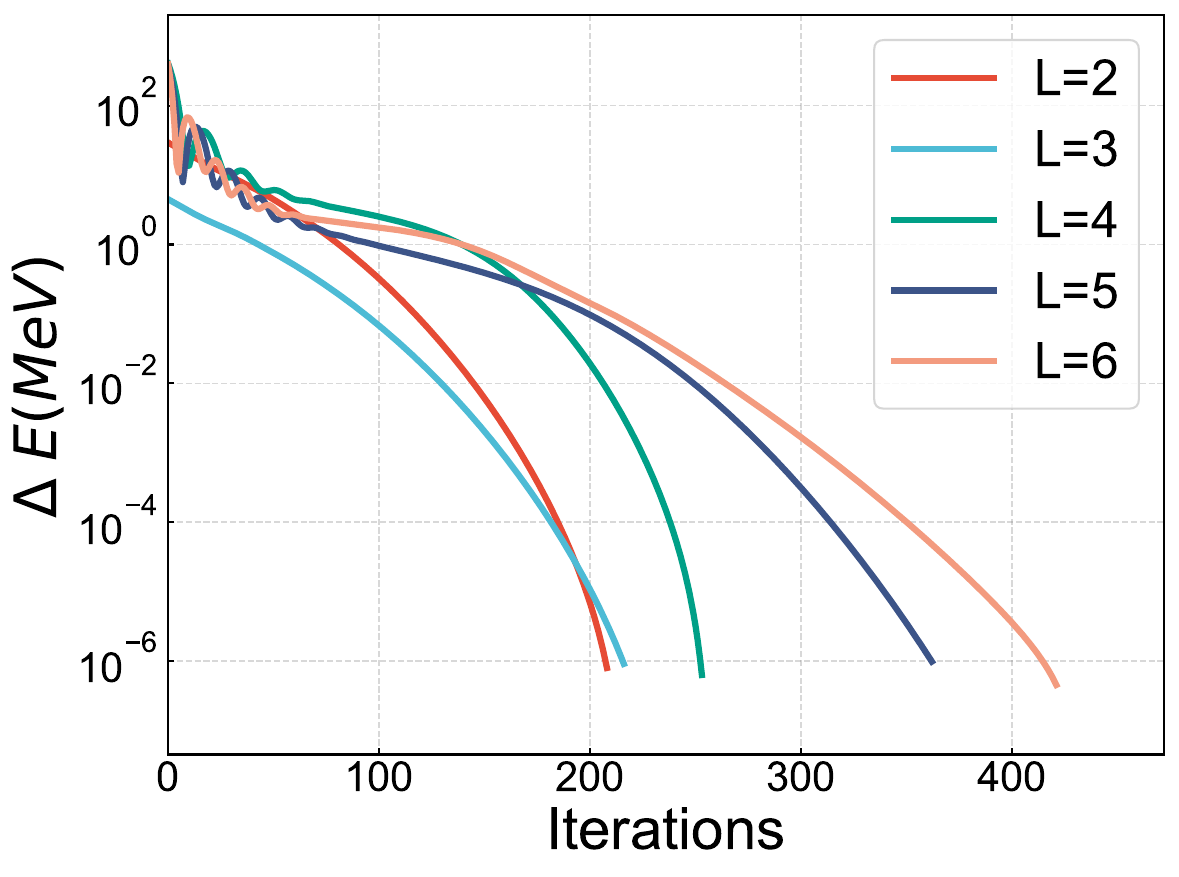} 
        \label{N2L}
    }
    \hfill 
    \subfloat[Triton($^3$H)]{
        \includegraphics[width=0.32\linewidth]{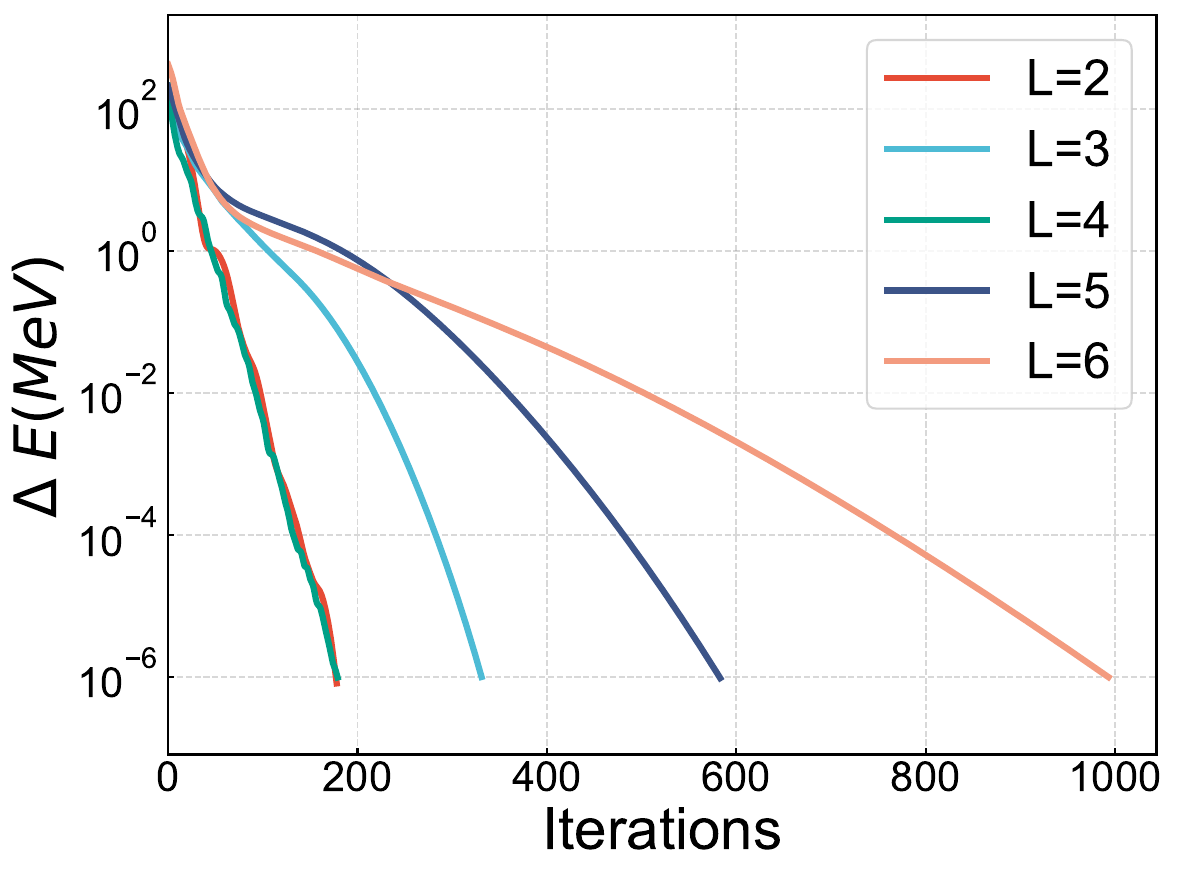} 
        \label{N3L}
    }
    \hfill 
    \subfloat[Helium-4($^4$He)]{
        \includegraphics[width=0.32\linewidth]{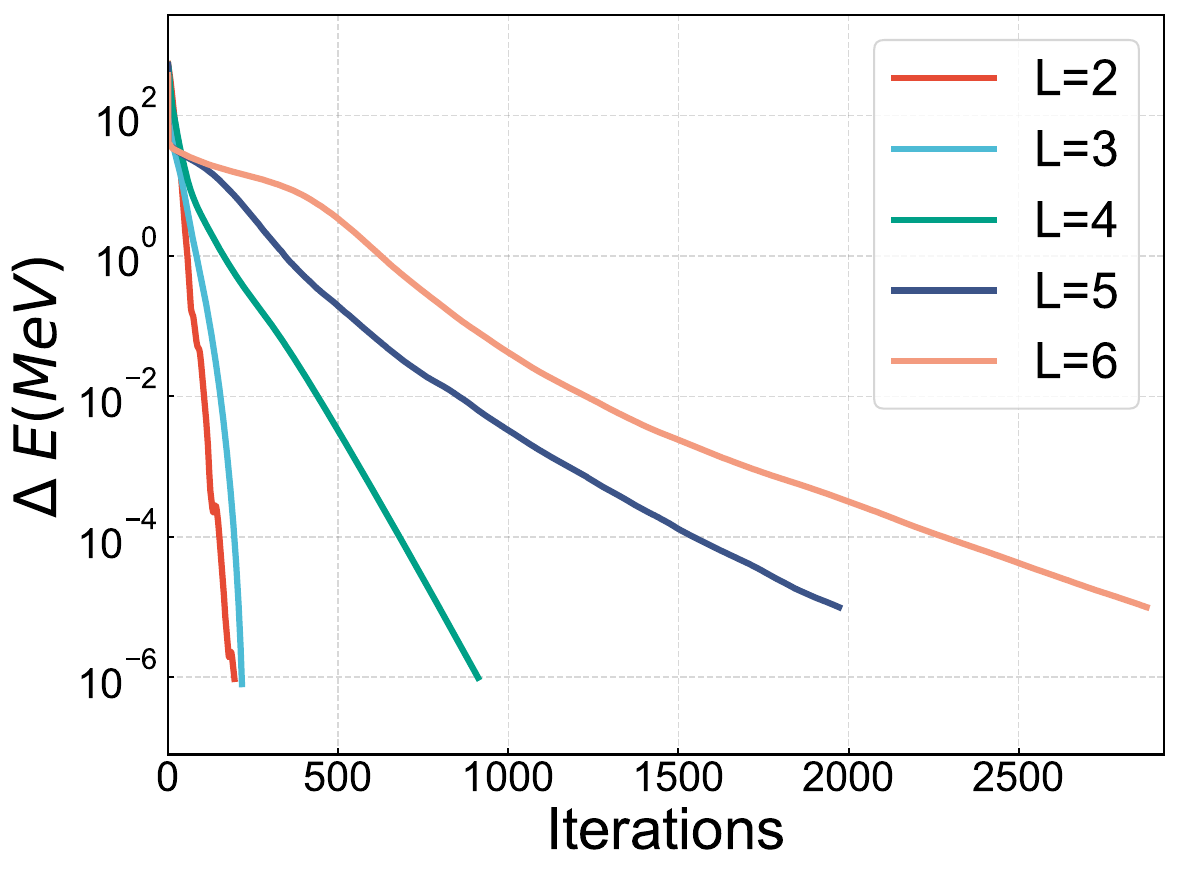} 
        \label{N4L}
    }

    \caption{Convergence profiles of the variational ground-state energy for light nuclei: (a) deuteron ($^2$H, $n=2$), (b) triton ($^3$H, $n=3$), and (c) helium-4 ($^4$He, $n=4$). The energy residual, defined as $\Delta E(\theta_t) = E(\theta_t) - E_{\text{0}}$ (where $E_{\text{0}}$ is the exact-diagonalization energy of the Hamiltonian at the same $L$), is plotted on a semi-logarithmic scale against the optimization iteration $t$ for various lattice dimensions $L$. 
    }
    \label{NL}
\end{figure*}

Figure~\ref{NL} shows the convergence behavior of the VQE for deuteron ($n=2$), triton ($n=3$), and helium-4 ($n=4$). We plot the energy residual $\Delta E$ as a function of iteration and set the stopping accuracy of $10^{-6}~\mathrm{MeV}$ in our simulations.

The number of iterations needed to reach this level depends on $n$ and $L$. When $n$ increases, convergence becomes slower. At fixed $n$, increasing $L$ also makes the optimization harder. This trend is consistent with the fact that the effective Hilbert space grows with $n$ and $L$, so the optimization has to search a larger parameter space.

\section{Conclusion}

In this work, we presented a proof-of-principle study of quantum computing for an effective nuclear lattice model. We formulated the Hamiltonian in second quantization with two-body and three-body contact interactions, and used the variational quantum eigensolver to compute ground-state energies. To address the resource limitations of near-term hardware, we systematically compared the Jordan-Wigner transformation and Gray code encoding. For the few-body systems investigated herein, the Gray code encoding, when applied in conjunction with symmetry reductions, yields a substantially diminished qubit count: it grows only logarithmically with $L^3$, in stark contrast to the cubic $\mathcal{O}(L^3)$ scaling inherent to the Jordan-Wigner scheme. 

We applied this framework to $^{2}\mathrm{H}$, $^{3}\mathrm{H}$, and $^{4}\mathrm{He}$ on finite lattices with $L=2$--$6$. The results show that finite-volume effects are strong at small $L$, but decrease as the lattice size increases, with the calculated ground-state energies approaching the corresponding experimental binding energies. We also observed that the VQE optimization becomes more difficult for larger particle numbers and lattice sizes, reflecting the growth of the effective Hilbert space.

The present study is restricted to classically emulated VQE calculations of few-body systems utilizing a relatively simple effective Hamiltonian with purely contact interactions. Nevertheless, these initial results provide a useful proof-of-principle starting point for future investigations. Algorithmic extensions may include more efficient encoding strategies, improved optimization schemes, and, ultimately, more efficient quantum algorithms. From a nuclear physics perspective, future progress will naturally entail expanding this framework to accommodate more complex and realistic nuclear lattice Hamiltonians, ultimately paving the way for the quantum simulation of medium-mass and heavy nuclei on scalable hardware.

\begin{acknowledgments}
This work is supported by NSAF ( No. U2330201 and No. U2330401),  National Natural Science Foundation of China with Grant No. 12275259, 12547105, Science Challenge Project (No. TZ2025012, No.TZ2025017) and Quantum Science and Technology-National Science and Technology Major Project (Grant No. 2023ZD0300200).
\end{acknowledgments}

\bibliography{main}

\end{document}